# Bibliometrics for Internet Media: Applying the *h*-Index to YouTube

Robert Hovden
School of Applied and Engineering Physics, Cornell University, Ithaca, NY 14853
rmh244@cornell.edu
(Dated: Jan 3, 2013)

The *h*-index can be a useful metric for evaluating a person's output of Internet media. Here we advocate and demonstrate adaption of the *h*-index and the *g*-index to the top video content creators on YouTube. The *h*-index for Internet video media is based on videos and their view counts. The index $h$ is defined as the number of videos with $\geq h \times 10^5$ views. The index $g$ is defined as the number of videos with $\geq g \times 10^5$ views on average. When compared to a video creator's total view-count, the *h*-index and *g*-index better capture both productivity and impact in a single metric.

**Introduction:**

A growing use of bibliometrics into the performance evaluation of academic authors has followed an increasing availability of citation indexes via digital database services such as *Web of Knowledge* or *Google Scholar*. In particular, the *h*-index has become widely popular and is now integrated into most scholarly databases (Bar-Ilan, 2007). The ease of calculation and ability to incorporate quantity and visibility of publications makes it an appealing metric (Hirsch, 2005). Following Hirsche's proposition of the *h*-index, deeper analysis and a slew of alternative bibliometric indices have been introduced (Acuna, Allesina, & Kording, 2012; Egghe, 2006a; Evans, 2008). The widespread adoption of these indices reflects their utility in quantifying scientific output (Bornmann & Daniel, 2007). However, these metrics need not be restricted to academic publication and can be readily adapted to other fields—including those that are not scholarly. In particular, popular Internet media websites—most notably, YouTube—act as publishers for content creators. These content creators span a wide spectrum of interests such as comedy, music, video blogging, science & education, news & politics, or technology. Evaluating the achievement of Internet media creators has become exceedingly relevant to consumers and investors of the nascent industry.

YouTube is currently the largest host to streaming video content on the Internet. Users upload videos to their channel, after which, the video is made accessible to the world for an indefinite period of time. Typically, video creators will regularly or sporadically upload new content to their channel. Each video receives one additional view count each time the video is watched through YouTube. Only limited data, such as the video view count for each video are made public. Most often the total view counts—the sum of view counts over all videos in a channel—is the metric for ranking the success of content creators. However, a channel's total views may be an inflated metric for success that is biased by a small number of 'big hit' videos. It therefore fails to measure the broad impact and productivity of a content creator. YouTube channels can also have subscribers—people that wish to receive notification of a channel's new content. The number of subscribers is also a valued metric for evaluating a channel's success. Obtaining more subscribers usually requires both impact and productivity, but is also convoluted with externalities such as user sentiment. Additionally, it is not a monotonic function as channel's subscribers can decrease or increase at any point in time.

Here popular bibliometric indices are adapted to best evaluate the performance of the top YouTube video creators. More specifically, we apply the *h*-index and the *g*-index to the content of each user or video "channel". Simply comparing a video channel's total views is a poor measure of a creator's total productivity—often inflated by a small number of viral videos. Analogous to academia, the *h*-index for YouTube overcomes this limitation by evaluating a content creator's productivity and impact in a single indicator. The *h*-index has become particularly popular and therefore was chosen to best demonstrate the applicability of bibliometrics to Internet media. The *g*-index, similar to the *h*-index, may have marginal benefits by giving more weight to content with high impact (Costas & Bordons, 2008; Egghe, 2006b). These bibliometrics, and others, can be applicable to any media content host that tracks viewership. Here, we focus on contributors to YouTube because of its dominant popularity and readily automated access to data via the YouTube API (see Appendix I, II).

A YouTuber—i.e. a video content creator—has an index *h* if *h* of his or her $N_v$ videos have at least $h \times 10^5$ views each and the other $(N_v - h)$ videos have $< h \times 10^5$ views each.

Similarly, a YouTuber has an index *g* if his or her $N_v$ videos have at least $g^2 \times 10^5$ views. Or alternatively stated, *g* is the number of highly viewed videos that have on average at least $g \times 10^5$ views.

**Methods:**

In the adaptation and application of the *h*-index to a YouTube channel, the view counts of each video determine its *h*-index. A high *h*-index requires both productivity (a large number of videos) and high impact (videos with many views). The *h*-index of academic publications is a monotonically increasing function, except in rare instances where papers are retracted. However, removal of content is more common with Internet-based resources like YouTube videos and a user's *h*-index could decrease if a sufficiently popular video is removed. The *h*-index for Internet media, as we define, is the number of videos N that have N×100,000 views or more. By this definition, we are making $10^5$ videos views analogous to 1 citation in academia. YouTube is acting as the publisher and a particular YouTube channel or user account can be viewed as the author. YouTube does not currently index multiple contributors—i.e. video credits—so multiple authors of a single video is not possible.

While the choice of $10^5$ views is somewhat arbitrary, it is an order of magnitude that produces *h*-index values of top YouTubers most consistent with the top academics. For strictly academic

YouTube videos, a lower threshold (perhaps $10^4$) may be a better choice to accommodate their relatively low view counts. In Hirsche's paper, he reports the mean and median *h*-index of Nobel Prize winning physicists (years 1975-2005) to be 41 and 35 respectively. Here, using a threshold of $10^5$ views, we find the top 25 YouTube *h*-indexes to have a mean and median of 56.7 and 55 respectively. The current highest YouTube *h*-index of 79 belongs to *smosh*, meaning his channel has 79 videos each with over 7.9 million views. His first video was only posted three years ago. The mean age of these top 25 YouTube channels is 4.20 years, a relatively short amount of time when compared to the life of most academic papers. Given the relative youth of Internet media, we can expect YouTube *h*-indexes to rise noticeably with age and increasing Internet viewership.

The *g*-index is similar to the *h*-index but looks at the average views of the top *g* videos. The *g*-index is the number of highly viewed videos that have on average at least $g \times 10^5$ views. This places more value on those channels with a few highly viral videos. The *g*-index will take on values no smaller and most often larger than the *h*-index.

**Results:**

Table 1 shows the ranking of YouTube channels based on their total channel views, *h*-index, *g*-index, and subscribers. Depending on the metric of choice—total views, *h*-index, *g*-index, or subscribers—different users will outperform others. What becomes immediately apparent is ranking based on total view counts is least consistent with the other index rankings. A shift from total view-count to adoption of the *h*-index in ranking top YouTube channels would change the top YouTube channel from *justinbiebervevo* to *smosh*. More dramatically, *nigahiga* is ranked 25th based on total views but soars to 3rd in *h*-index—a value that more closely matches his 2nd place rank in subscribers. Similarly, *jennamarbles* does not fall in the top 25 for total views but is the 7th highest *h*-index and the 5th highest number of subscribers. When looking at total views, music video based channels (e.g. *justinbiebervevo*, *rihannavevo*, *ladygagavevo, officialpsy*) with a handful of heavily watched videos have a strong presence. However in the *h*-index rankings, video blogs and mini-shows (e.g. *raywilliamjohnson*, *smosh*, *realannoyingorange*) with more prolific video output lead the rankings. The *g*-index gives more value to those channels with heavily watched videos and has a ranking somewhere between the *h*-index and total view counts. For this reason, many music video channels still appear among the top 25 *g*-indexes.

Based upon the 50 most subscribed channels, the correlation of the YouTube *h*-index, *g*-index, and total views to a channel's subscribers, we find Pearson correlation coefficients of 0.68, 0.47, and 0.38 with p-values of $1.8 \times 10^{-8}$, $4.0 \times 10^{-4}$, and $5.0 \times 10^{-3}$ respectively. This indicates that the *h*-index has the strongest correlation with the number of subscribers when considering top YouTube channels.

Hirsch also proposed normalizing the h-index by the time since first publication. This can be applied to YouTube channels through dividing the YouTube h-Index by the number of active years of the channel (the oldest published video). Among the top YouTube channels in Table 1, the top five channels based on a normalized h-index rank are: *ultrarecords* 22.20, *raywilliamjohnson* 20.92, *muyap* 16.87, *JennaMarbles* 16.39, and *collegehumor* 15.47. A normalized h-Index could provide a metric favoring content with longevity over short-lived trendy channels.

In academic publications, the *h*-index is criticized for its poor ability in comparing scholars from different fields with different citation behavior (Barendse, 2007). Thus, the h-index best compares authors within a particular area of research. Inherent differences between the audiences of different YouTube video types will also affect the metrics of performance in digital media. For example, a video debating the subtleties of a particular economic policy might have a much smaller audience than a pop music video. Table 2 shows the rankings of four different YouTube Channel types. The Top "Reporters" channels have lower view counts and *h*-indices when compared to "Comedians" or "Musicians". This suggests *h*-indexes are best used to compare YouTube channels in a related field. Currently there are only nine channel types permitted by YouTube and, for most, there is little oversight to how a user categorizes their channel—which may not be categorized at all. However, Table 2 still provides insight to the differences between content types on YouTube. Similarly, cultural, geographic, and language differences would be expected to also influence the performance of a YouTube channel.

**Discussion:**

The proposed bibliometrics for YouTube have demonstrated utility, however there are differences to consider when drawing a direct analogy between the h-index of academic publication and that of Internet media. A citation implies a manuscript influenced future published work whereas a video's view count signifies that it has piqued the interest of someone. An immerging field of alternative metrics (Banks & Dellavalle, 2008; Priem & Hemminger, 2010) has investigated the use of downloads and view-counts in quantifying the productivity and impact of scholars (Priem, Piwowar, & Hemminger, 2011; Shuai, Pepe, & Bollen, 2012). Strong correlations have been found between the citation and download impact of peer-reviewed articles (Brody, Harnad, & Carr, 2006). This suggests YouTube view counts—with near similarity to download counts—provides a reasonable approach to an alternative $h$-index quantification where citations are not present.

While literature citations comes with known difficulties (M. H. MacRoberts & MacRoberts, 1996), so does the performance analysis of Internet media. View counts are susceptible fraudulent inflation, limited categorization within YouTube provides challenges in comparing videos from different fields, and acquisition of data from hosting companies of Internet media is usually limited or impossible. For mass media with widespread appeal, the h-index and other bibliometrics of Internet media provide insight. However, the adoption of an h-index—based on viewcount—for evaluating the academic performance of scholarly videos is likely premature.

**Conclusion:**

Supreme court justice Alito forecasted in January 2012 the inevitable transition from traditional television media into Internet media when he stated, "broadcast TV is living on borrowed time, it is not going to be long before it goes the way of vinyl records and 8 track tapes" (FCC v. Fox Television). With the heralded shift to Internet media, the importance in quantitatively evaluating the success of Internet content creators—e.g. YouTube users—becomes an increasingly relevant task. In just the last few years, new metrics have been proposed to better quantify an academic author's performance. Their utility has been rigorously investigated and is best suggested by their widespread adoption. We can expect the same from analogous YouTube metrics—such as the $h$- or $g$- index. More complex metrics tailored to Internet media that integrate ratings, comments, as well as subscribers, could have advantages over traditional bibliometrics. However, the $h$-index provides a stable metric for achievement and requires only minimal counting to calculate. There is solace in the simplicity, universality, and popularity of the $h$-index when applying it to YouTube.

**Acknowledgements:**

RH thanks Kara M Church for proof reading and feedback on the manuscript, Dustin L Walsh for informative discussions on YouTube culture., and the anonymous referees who helped improve content.  RH also thanks Michael at VidStatsX for informative discussions regarding the YouTube data API and its shortcomings.

| Bibliometrics Rankings of Top YouTube Channels | | | | | | | |
|---|---|---|---|---|---|---|---|
| Total Views (millions) | | h-index | | g-index | | subscribers (thousands) | |
| 3280 | JustinBieberVEVO | 79 | smosh | 141 | AtlanticVideos | 6141 | raywilliamjohnson |
| 3175 | RihannaVEVO | 77 | RayWilliamJohnson | 130 | UltraRecords | 6024 | nigahiga |
| 2210 | AtlanticVideos | 70 | nigahiga | 128 | FueledByRamen | 5844 | smosh |
| 2184 | smosh | 69 | realannoyingorange | 118 | smosh | 5123 | machinima |
| 2177 | EminemVEVO | 64 | UltraRecords | 115 | realannoyingorange | 4706 | jennamarbles |
| 2141 | RayWilliamJohnson | 61 | nqtv | 110 | barelypolitical | 3763 | freddiew |
| 2131 | LadyGagaVEVO | 61 | JennaMarbles | 109 | nigahiga | 3222 | rihannavevo |
| 1991 | UltraRecords | 59 | MondoMedia | 104 | linkinparktv | 3123 | collegehumor |
| 1834 | shakiraVEVO | 58 | AtlanticVideos | 101 | kontor | 2982 | shanedawsontv |
| 1726 | FueledByRamen | 58 | Fred | 99 | nqtv | 2920 | fpsrussia |
| 1668 | beyonceVEVO | 57 | huluDotCom | 97 | Fred | 2861 | epicmealtime |
| 1608 | officialpsy | 56 | barelypolitical | 96 | SpinninRec | 2715 | pewdiepie |
| 1553 | barelypolitical | 55 | muyap | 96 | huluDotCom | 2690 | bluexephos |
| 1498 | hollywoodrecords | 55 | freddiew | 94 | MondoMedia | 2573 | realannoyingorange |
| 1487 | realannoyingorange | 54 | kontor | 93 | RovioMobile | 2515 | thelonelyisland |
| 1445 | BlackEyedPeasVEVO | 54 | BritainsGotTalent09 | 93 | JennaMarbles | 2499 | tobuscus |
| 1439 | ChrisBrownVEVO | 54 | boyceavenue | 92 | BritainsGotTalent09 | 2500 | kevjumba |
| 1429 | muyap | 50 | machinima | 92 | TheOfficialSkrillex | 2460 | werevertumorro |
| 1423 | machinima | 48 | FueledByRamen | 92 | davidguetta | 2417 | riotgamesinc |
| 1421 | JenniferLopezVEVO | 48 | TheXFactorUK | 90 | Flowgo | 2360 | michellephan |
| 1411 | kontor | 47 | beyonceVEVO | 88 | sment | 2333 | roosterteeth |
| 1384 | PitbullVEVO | 47 | ShaneDawsonTV | 88 | RayWilliamJohnson | 2325 | onedirectionvevo |
| 1376 | KatyPerryVEVO | 46 | collegehumor | 86 | warnerbrosrecords | 2292 | justinbiebervevo |
| 1354 | MondoMedia | 46 | warnerbrosrecords | 84 | daneboe | 2253 | sxephil |
| 1336 | nigahiga | 44 | SpinninRec | 83 | thelonelyisland | 2143 | barelypolitical |

TABLE 1 | Rankings of top 25 YouTube channels based on four metrics: total views, *h*-index, *g*-index, and subscribers. Rankings shift between metrics chosen. The *h*-index and *g*-index are popular in academia for their ability to quantify impact and productivity. Total views is often influenced too strongly by a few popular videos. The *h*-index is simple to calculate and correlates well with subscribers. Data collected from published videos on Jan 3, 2013.

| Top Rankings for Different YouTube Channel Types | | | | | |
| --- | --- | --- | --- | --- | --- |
| Comedians | | | Musicians | | |
|  | $h$-index | total-views |  | $h$-index | total-views |
| 'smosh' | 79 | 2153 | 'UltraRecords' | 64 | 1990 |
| 'RayWilliamJohnson' | 77 | 2140 | 'AtlanticVideos' | 58 | 2209 |
| 'nigahiga' | 70 | 1304 | 'boyceavenue' | 54 | 808 |
| 'realannoyingorange' | 69 | 1486 | 'kontor' | 54 | 1410 |
| 'nqtv' | 61 | 1021 | 'FueledByRamen' | 48 | 1725 |
| 'Fred' | 58 | 949 | 'beyonceVEVO' | 47 | 1667 |
| 'collegehumor' | 46 | 1136 | 'UKFDubstep' | 42 | 920 |
| 'AdamThomasMoran' | 44 | 385 | 'RihannaVEVO' | 41 | 3172 |
| 'TheEllenShow' | 41 | 1052 | 'shakiraVEVO' | 39 | 1833 |
| 'werevertumorro' | 40 | 732 | 'linkinparktv' | 37 | 1096 |
| Gurus | | | Reporters | | |
|  | $h$-index | total-views |  | $h$-index | total-views |
| 'FPSRussia' | 40 | 490 | 'AssociatedPress' | 31 | 609 |
| 'MichellePhan' | 38 | 626 | 'Matroix' | 19 | 262 |
| 'kipkay' | 33 | 378 | 'ABCNews' | 18 | 290 |
| 'Howcast' | 25 | 524 | 'www16barsde' | 16 | 145 |
| 'expertvillage' | 24 | 517 | 'JuliensBlog' | 15 | 116 |
| 'bubzbeauty' | 21 | 256 | 'TMZ' | 14 | 127 |
| 'HouseholdHacker' | 20 | 201 | 'IshatOnU' | 14 | 110 |
| 'CaptainSparklez' | 19 | 591 | 'CTFxC' | 12 | 186 |
| 'TobyGames' | 19 | 423 | 'FUNKER530' | 12 | 105 |
| 'dope2111' | 18 | 143 | 'scoutthedoggie' | 12 | 109 |

TABLE 2 | Comparison of top 10 YouTube channels of four different "user types" permitted. Here, The $h$-index is calculated from the view-count of the videos in a channel (based on $10^5$ views). Variation in the h-index between categorical types can be seen. Top "Reporters" channels have lower view counts and $h$-indices when compared to "Comedians" or "Musicians". Note: Not all YouTube channels are categorized and present in a "user type" YouTube API Channel Feed. Data collected from published videos on Jan 3, 2013.

**Appendix I: Retrieving The Most Subscribed / Viewed YouTube Channels**

A list of the most subscribed channels can be obtained from a channel feed in the YouTube API v2.0. However, due to current inaccuracies with the YouTube channel feed (bug report #3748) the results were cross-referenced with the data from VidStatsX.com. It was found that a couple top channels were missing from the YouTube most subscribed channel feed. Below is a Python script that retrieves and parses the 100 most subscribed channels from the YouTube API:

```
#Written for Python v. 2.7.1, Feedparser v. 5.1.3
import feedparser
print '\n--Retrieving Most Subscribed Channels--\n'
ftop100     = open('top100.txt','w')
for start in range(1, 101, 50):
   uri = 'http://gdata.youtube.com/feeds/api/channelstandardfeeds/most_subscribed?start-index=' + str(start) + '&time=all_time&&max-results=50&v=2'
   feed = feedparser.parse(uri)
   for post in feed.entries:
      print post.author
      ftop100.write( post.author + '\n' )
ftop100.close()
```

Changing the uri from "`most_subscribed`" to "`most_viewed`" will retrieve the most viewed YouTube channels. Similarly, adding a channel type suffix to either will retrieve top channels of a particular type (e.g. "`most_subscribed_Comedians`"). Currently nine channel types are permitted: Comedians, Directors, Gurus, Musicians, Non-Profit, Partners, Politicians, Reporters, Sponsors.

**Appendix II: Retrieving A YouTube Channel's Videos And Their Views**

Python scripting and the The Google Data API can retrieve information on every video in a YouTube channel. The script below retrieves the view count of every video over a list of YouTube Channels. A text file containing a list of every user with descending list of every video's view. Also a list of the users and their respective subscriber count is created. Note that a sleep statement to pause the program can prevent over-accessing Google Data services.

```
#Written for Python v. 2.7.1, Google Data API 2.0
import gdata.youtube
```

```python
import gdata.youtube.service
import time

yt_service = gdata.youtube.service.YouTubeService()
yt_service.ssl = True

def GetAndWriteEntryStats(uri,username,fviews,fratings,fdurations):
  yt_service = gdata.youtube.service.YouTubeService()
  feed = yt_service.GetYouTubeVideoFeed(uri)
  for entry in feed.entry:
    WriteEntryStats(entry,username,fviews,fratings,fdurations)

def WriteEntryStats(entry,username,fviews,fratings,fdurations):
  try:
    fviews.write(entry.statistics.view_count + '\t')
  except:
    fviews.write('na' + '\t')

#START MAIN HERE
print '\n--Running Youtube View Count Analyzer--\n'

with open('top100.txt') as fusernames:
    usernames = fusernames.read().splitlines()
fusernames.close()

fviews       = open('views.txt','w')
fsubscribers = open('subscribers.txt','w')

for username in usernames:
  print '- - - - - ' + username + ' - - - - - -'
  fviews.write(username + '\t')
  for start in range(1, 501, 50):
    uri = 'http://gdata.youtube.com/feeds/api/users/' + username + '/uploads?start-index=' + str(start) + '&max-results=50&orderby=viewCount&racy=include'
    GetAndWriteEntryStats(uri,username,fviews,fratings,fdurations)
  fviews.write('\n')
  uri = 'http://gdata.youtube.com/feeds/api/users/' + username
  user_entry = yt_service.GetYouTubeUserEntry(uri)
  fsubscribers.write(user_entry.username.text + '\t')
  fsubscribers.write(user_entry.statistics.subscriber_count + '\n')

  time.sleep(6)

print '\n'
```